**ORIGINAL RESEARCH**

# Standardizing Paediatric Clinical Data: The Development of the conect4children (c4c) Cross Cutting Paediatric Data Dictionary

Anando Sen*, Victoria Hedley*, John Owen†, Ronald Cornet‡, Dipak Kalra§, Corinna Engel‖, Avril Palmeri*, Joanne Lee*, Jeane Christophe Roze¶, Joseph Standing**, Adilia Warris††, Claudia Pansieri‡‡, Rebecca Leary*, Mark Turner§§ and Volker Straub*

**Introduction:** Standardization of data items collected in paediatric clinical trials is an important but challenging issue. The CDISC data standards are well understood by the pharmaceutical industry but lack the implementation of some paediatric-specific concepts. When a paediatric concept is absent within CDISC standards, pharmaceutical companies and research institutions take multiple approaches in the collection of paediatric data, leading to different implementations of standards and potentially limited utility for reuse.
**Objective:** To overcome these challenges, the conect4children consortium has developed a cross-cutting paediatric data dictionary (CCPDD).
**Methods:** The dictionary was built over three phases: scoping (including a survey sent out to ten industrial and 34 academic partners to gauge interest), creation of a longlist and consensus building for the final set of terms. The dictionary was finalized during a workshop with attendees from academia, hospitals, industry and CDISC. The attendees held detailed discussions on each data item and participated in the final vote on the inclusion of the item in the CCPDD.
**Results:** Nine industrial and all 34 academic partners responded to the survey, which showed overall interest in the development of the CCPDD. Following the final vote on 27 data items, three were rejected, six were deferred to the next version and a final opinion was sought from CDISC. The first version of the CCPDD with 25 data items was released in August 2019.
**Discussion and Conclusion:** The continued use of the dictionary has the potential to ensure the collection of standardized data that is interoperable and can later be pooled and reused for other applications. The dictionary is already being used for case report form creation in three clinical trials. The CCPDD will also serve as one of the inputs to the Paediatric User Guide, which is being developed by CDISC.

## Introduction

Paediatric clinical trials face substantial challenges due to age-specific formulations and dosages, vulnerable and often small populations affected by specific diseases, lengthy regulatory challenges, ethical considerations, and hesitance on the part of parents or guardians to enrol children.[1–6] Pooling clinical data from different sources can lead to the generation of large cohorts for post-hoc studies as well as meaningful comparisons between studies.[7,8] However, interoperability (both semantic and syntactic) between data obtained from different sources for different purposes remains an issue.[9,10] For example, body temperature measured orally, at the forehead or under the arm cannot be treated as the same measurement. Hence, standardization of paediatric data is an important scientific challenge. A grant-funded Paediatric working

* Newcastle University, Newcastle upon Tyne, UK
† Clinical Data Interchange Standards Consortium (CDISC) Europe Foundation, Brussels, BE
‡ Amsterdam University Medical Centers, Amsterdam, NL
§ The European Institute for Innovation through Health Data, Gent, BE
‖ University Children's Hospital, Tubingen, DE
¶ Nantes University, Nantes, France
** University College London, London, UK
†† University of Exeter, Exeter, UK
‡‡ Consorzio per Valutazioni Biologiche e Farmacologiche, Pavia, IT
§§ University of Liverpool, Liverpool, UK

Corresponding author: Rebecca Leary (becca.leary@newcastle.ac.uk)

group[11] within Health Level Seven (HL7) set out to define a set of data elements to support both care and research but lacked ongoing funding, resulting in the work not reaching the ballot. Further, several National Institute of Health (NIH)-funded paediatric clinical trial networks are active but are not pursuing development of data standards.[12,13]

There is a lack of consensus about how to standardize the collection of many of the paediatric specific data items in paediatric clinical studies. Standards developed by the Clinical Data Interchange Standards Consortium (CDISC) are well understood by the pharmaceutical industry. Their value is reflected in the requirement for their use in electronic regulatory submissions by the US Food and Drug Administration (FDA) and the Japanese Pharmaceutical and Medical Devices Agency (PMDA), and preference for their use in submissions to China's National Medical Products Administration.[14–16] However, CDISC standards currently have two main drawbacks: 1) while CDISC foundational standards can be leveraged, they lack specific paediatric concepts required by sponsors of paediatric clinical trials,[17,18] and 2) CDISC standards are less understood in academic clinical research.[19]

Where no CDISC standards exist (or implementation examples for existing standards are not provided), pharmaceutical companies take several different approaches. These include using company-specific standards (often following the general CDISC structure), with the option to directly engage with CDISC to include these in the CDISC standards. Hence, each sponsor of a trial essentially creates data dictionaries that may contain study-specific or institution-specific standards.[20] These data dictionaries are proprietary in nature and are generally not shared. This lack of standardization results in heterogeneity in the way the data is represented, making interoperability (i.e., pooling and sharing in a meaningful way) very difficult and in some cases impossible. In addition, a lack of data standards leads to a lower efficiency of trial execution as each organisation must start from scratch when developing new data collection tools and standards.

These issues are concerning as, guided by the findable, accessible, interoperable, reusable (FAIR) principles,[21] clinical research is increasingly moving towards an era of data reuse and interoperability. The FAIR principles for data management, published in 2016, were based on the ability of computer systems to use data for research with minimal human intervention. In addition to paediatrics, the need for standardized data is well recognized in the field of rare diseases (according to the European Commission, serious diseases with a prevalence lower than 1 in 2000 that require special combined efforts to address them), where speed and accuracy of diagnosis, knowledge about disease trajectory and potential treatment options can all be improved through knowledge obtained from prior data. Rare diseases are of importance to paediatrics as most rare diseases affect children or have an exclusive paediatric onset. Given the scarcity of rare diseases data, pooling and sharing of data is critical.[22,23]

## Background

Motivated by the aforementioned issues, the conect4children (c4c) consortium was established in 2018. The main objective for c4c is to address the multitude of barriers facing the development and delivery of effective paediatric clinical trials.[24] c4c is a time-limited public–private consortium funded by the Innovative Medicines Initiative (IMI). IMI projects are co-financed by the European Commission and by the pharmaceutical industry through its European association, European Federation of Pharmaceutical Industries and Associations (EFPIA). The research areas within c4c are organized into eight work packages. Work package five is dedicated to data standardization, whose goals are summarized in **Box 1**.

One of the first steps for the data standardization work package was the development of a cross-cutting paediatric data dictionary (CCPDD). Given its intended use as an input to the CDISC Paediatrics User Guide (PUG), CDISC was fully involved in the development of the dictionary. This ensured the dictionary would be ready for use immediately upon release for standardizing case report forms (CRFs) in paediatric clinical trials. The focus of this dictionary was on cross-cutting (or disease-independent) data items that are routinely collected for clinical research. This paper describes the rigorous consensus building methods that were used to create the dictionary. Discussions about contentious data items are included. Finally, the improvements to be made in the next iteration of the dictionary and extension to the PUG are discussed.

---

**Box 1:** Summary of the conect4children Work Package 5 vision.

**Summary of conect4children Work Package 5 goals**

- Increase in standardized clinical trial data and metadata will result in an increase in scientific knowledge about paediatric diseases. This knowledge will enable the development of safer medicines for children.
- High quality data-driven knowledge has the potential to develop better, more efficient paediatric clinical trial designs. Trials will be able to harness the power of standardized data to employ more innovative methods for paediatric medicine development (e.g., repurposing the same data for multiple applications, extrapolation of findings from previously acquired data).
- The burden on children and their families, who had to take part in multiple clinical trials (often duplicated or unnecessary) will be reduced.[25,26]
- Science based on FAIR data will increase its impact and will improve societal acceptance of research in children.[27]
- Standardized data could enable meta-analyses based on individual patient-level instead of aggregated data.

## Materials and Methods
The first iteration of the c4c CCPDD was developed over about 16 months (May 2018 to August 2019) and included three major phases: preparation (scoping), creation of the longlist of data items and consensus building.

### Phase 1 – Preparation (scoping)
*c4c survey*
To gauge interest among industry and academic partners in developing a CCPDD for standardizing CRFs, a short survey was developed by the Newcastle University c4c team. The majority of questions had multiple-choice answers, while some questions had an option to provide written feedback. This survey was created online using Lime survey (https://www.limesurvey.org/) and sent to c4c's ten pharmaceutical industry partners. A similar survey was also sent to the 34 academic and non-EFPIA beneficiaries. A blank copy of the survey is included as a supplementary document. It must be noted that some questions in the survey went beyond the scope of the CCPDD.

*Establishment of the CCPDD working group*
Following positive feedback from the survey (see Results), a small working group (WG) of 16 representatives comprised of paediatricians, academics, pharma, data standards experts and data scientists was established. This group took on responsibility for all strategic decision making related to the c4c CCPDD.

*Defining scope and purpose*
Before identifying data items to be included in the dictionary, its purpose and scope was defined by the WG. For practical reasons and to maximise the potential for reuse and added value, the WG agreed that this data dictionary should focus on specifying data items that would commonly be collected across multiple paediatric clinical trials and across multiple disease areas. It was also agreed to focus on the data items that might be collected in specific ways for children, rather than data items that would be captured similarly, universally, across all ages. It was decided that the dictionary would not mandate the method used for collecting a data item (e.g., it does not stipulate that blood pressure should be taken sitting down). However, it would strongly recommend the capture qualifiers on CRFs. These may be method qualifiers (e.g., blood pressure sitting down, lying down, standing up, etc.), time qualifiers (e.g., morning, afternoon, etc.) or any other measurement-related qualifier.

The following criteria for data item selection were agreed by the CCPDD WG and the c4c project leadership team: (1) all data items in the CCPDD must be relevant to paediatrics; (2) data items must be cross-cutting in nature, meaning they could not be disease-specific; (3) data items should be commonly collected in paediatric clinical trials, which could include data items that are relevant for one age group (i.e., puberty-related items for adolescents or items related to birth for neonates) but are not collected in trials in other age groups; (4) data items commonly collected in clinical practice, but unused within clinical trials, would be considered out of scope; and (5) where they exist, CDISC standards for data items would be referenced.

### Phase 2: Generating the longlist of data items
The generation of the longlist was an informal process where terms were collected from different sources listed below. The final decision on the inclusion of a particular term in the CCPDD was taken in Phase 3.

1. CRFs and data dictionaries provided by members of the c4c consortium – The CRFs were reviewed, and any disease-independent data items were noted. Some examples of the CRFs and dictionaries used included Bayer Global Standard CRF page and Haemophilia Standard CRF page, Novartis Tanner staging and vital sign CRFs, and TREAT-NMD Global SMA/DMD registry data dictionaries for baseline report, lab data, physiotherapy and visits.
2. Data items from publicly available sources – Keyword searches were performed on clinicalstudydata-request.com, yoda.yale.edu and vivli.org. Keywords included paediatrics, neonates, neonatal, adolescent, child, infant, children, babies, new-born, and puberty. The advanced search feature in clinicaltrials.gov was used to search paediatric studies (ages birth to 17). The text in the trial description, trial arms, eligibility criteria and outcome measures were scanned, and disease agnostic data items were noted.
3. Suggested data items from industry and academic partners – For some partners it was not possible to share CRFs or data dictionaries with c4c due to proprietary restrictions. Instead, some participants provided lists of data items for consideration. These came from real paediatric trials and studies from within their intuitions or clinical trial units.
4. Literature search on PubMed – A literature search was carried out on PubMed. Papers with results from paediatric studies were analysed. Any cross-cutting data items were noted for addition to the longlist. Data items that were disease-specific in nature were not included.

### Phase 3 – Consensus building
The longlist of items was reviewed by members of the WG and further revised at the first of two workshops held in Lisbon, Portugal in April 2019. The major task during this workshop was deciding which data items were cross cutting. Generally, a cross-cutting term was one that occurred in 80% of the CRFs. However, implementation of this definition was not straight-forward due to 1) the presence of similar but non-synonymous terms (e.g., sex vs phenotypic sex) and 2) sources other than CRFs being used to generate the longlist (e.g., PubMed, public sources, etc.). It was also decided to follow CDISC standards to the maximum extent possible. Over the course of the workshop and a series of calls after, the longlist was progressively shortened by consensus. Wherever consensus could not be reached, the term was carried forward to the next stage.

Some additional decisions were also taken for the next steps. The dictionary that had been split by age (into neonates, children, and adolescents) up to this point was instead split into data item categories: demographics, vital signs, pubertal status, and others. The large amount of material identified for consideration by the companies in phase 2 prevented consideration of each identified source of content. To lessen the likelihood of overlooking important concepts, companies were recontacted in phase 3 and asked to review the longlist for omitted cross-cutting concepts. Any suggested terms were added to the list. Another key action from this workshop was to accompany each item of the CCPDD with guidance about its usage, which could help c4c study teams improve the harmonization of all the data collected in their studies. However, it was decided that adherence to the guidance would not be mandatory. A follow-up workshop dedicated to finalizing the CCPDD was planned for June 2019.

The second workshop was held at Newcastle, United Kingdom. In addition to academic and industry attendees, this workshop also included representatives from CDISC as one of the targets was to finalize the CCPDD as an input to CDISC PUG. In total there were 25 attendees (19 in person and 6 virtually). In the first part of the workshop, participants at the meeting were provided with the current list of data items, which had already been scrutinised several times. This was a more formal scrutiny where workshop attendees were asked to discuss four aspects of the data items. It was anticipated that there may be several cross-cutting items yet to be modelled by CDISC or needing significant work to make their existing definitions appropriate for paediatric use. The four aspects scrutinized were

1. **Scope**: Is the item genuinely cross-cutting and is it routinely collected in paediatric trials?
2. **Units:** What are the appropriate units for the item (if applicable)?
3. **Qualifiers:** What timing or method qualifiers are needed (for example, position of patient for blood pressure measurements, method used for temperature collection, exact timing of a test result with respect to fasting)?
4. **Value ranges:** What are the appropriate value ranges for the item?

After the discussion on the first data item (head circumference), it was agreed that while defining extreme minimum and maximum ranges could be useful for data quality, it would be important to then consider what action might be triggered by an entry outside of this range. For example, if a study data system blocked such an entry from being entered. Because this would require further consensus, it was agreed to defer the inclusion of value ranges to a future iteration of the CCPDD. Hence, value ranges for the subsequent items were not considered. During the discussion, the participants could choose to replace an item by a similar item through unanimous consent.

The final decision was confirmed through a vote using the interactive voting software Vevox (https://www.vevox.com/live-voting-app). The software presented the percentage of respondents who agreed and disagreed on inclusion. The virtual attendees could also take part in this vote. A data item replaced by a similar item was not voted upon as the replacement process itself involved unanimous consent.

Following the vote, the participants had the option of deferring data items to the second iteration of the CCPDD if further consensus was required on qualifiers or guidance. The data items that had generated discussion, been replaced or been deferred were referred to CDISC for a final evaluation. Following feedback from CDISC, the data dictionary was finalized and released. The entire workflow on the development of the CCPDD is shown in **Figure 1**.

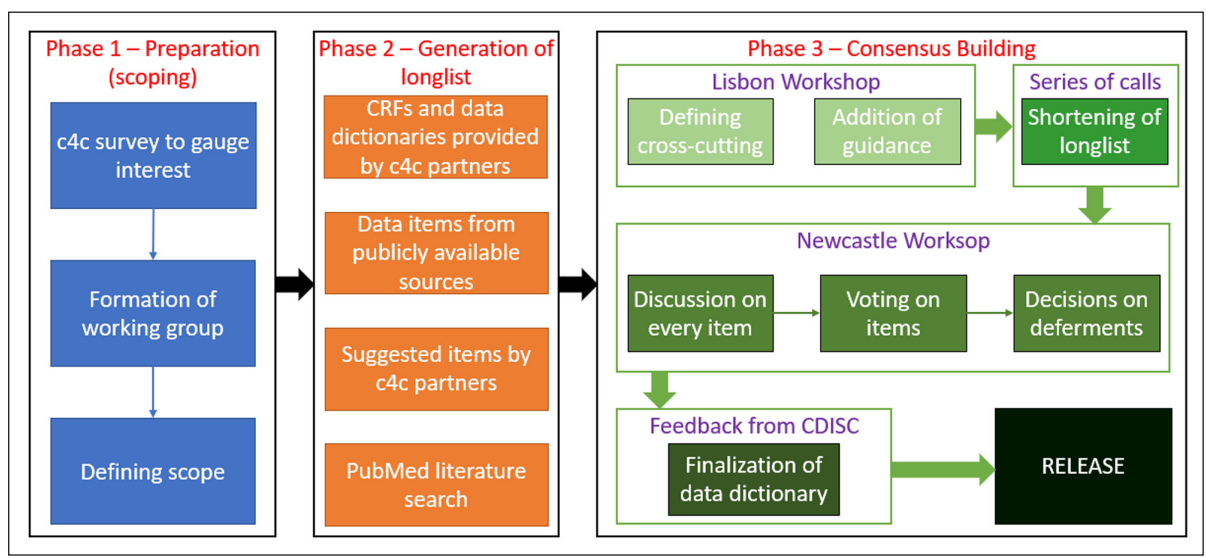

**Figure 1:** The sequence of events in the development of the cross-cutting paediatric data dictionary. The boxes are colour-coded as blue (phase 1), orange (phase 2) and shades of green (phase 3). The sequence of events is denoted by arrows. Textboxes without arrows within a larger box took place concurrently.

## Results

### Survey results

The industry survey had a 90% response rate (9/10). The survey indicated that all companies had standard operating procedures for CRF creation. Six of the nine companies use document architecture-based structuring standards for CRF creation (e.g., CDISC). Document architecture-based standards have a hierarchical structure that can be used to encode clinical documents. Eight of the nine companies believed it would be beneficial to establish a core set of cross-cutting data items that most paediatric clinical trials collect. All companies agreed that CRF-captured data must be interoperable. In fact, seven companies practiced pooling data from different trials. All but one company perceived a lack of standardization and harmonization in the way data items are recorded in paediatric trials as important to address.

The results from the academia and non-EFPIA beneficiaries were similar. There was a 100% (34/34) response rate. About 82% (28/34) of the respondents agreed it would be beneficial to have a core set of data items for paediatric trials, while one respondent did not agree. A vast majority of respondents (85%; 29/34) agreed that CRF-captured data must be interoperable (two answered "not very important" and three chose "no answer"). Almost 95% (32/34) of the respondents perceived the lack of standardization and harmonization in paediatric data storage as important to address (two chose "no answer").

The surveys suggested an overall interest in the development of a CCPDD, the main benefit of which would be standardization of data items and harmonization across institutions.

### Dictionary development

There were numerous iterations of the data dictionary over the 16 months. As the project unfolded, the final workshop occurred at Newcastle University, where final decisions were reached. Here, we present the summary of discussions surrounding a few of the contested data items. Because the discussion on head circumference led to value ranges being excluded (and qualifiers being optional in some cases) from further data items, we present the consequent discussion fully in **Box 2**.

Body mass index (BMI) had been added and removed from previous versions of the data dictionary several times, demonstrating lack of consensus. This item was finally removed from the data dictionary due to 1) normal ranges for BMI being based on specific populations and hence are not standardizable and 2) many clinicians contesting whether BMI was routinely collected across trials. It was unanimously agreed to include body surface area (BSA) instead, which is routinely collected in clinical trials as it is often used for dosing. A discussion took place on the use of the terms gender, sex, genotypic/phenotypic sex, but while it was decided to use the current CDISC definition for sex, the term to be used for the CCPDD was not confirmed. Sex and phenotypic sex were debated, and the final decision was deferred for further discussion during the development of the CDISC PUG. The results from the vote are detailed in **Table 1**.

---

**Box 2:** Summary of the discussion on head circumference that led to value ranges not being considered for future data items.

**Head circumference**

Current CDISC definition: A circumferential measurement of the head at the widest point.

In scope: The attendees agreed that this is a cross-cutting data item, particularly measured in neonatal trials. It was agreed that the measurement method should always be recorded, as c4c was not mandating a method. Any abnormality in the head circumference would need a separate data item but that item would not be cross cutting.

Units: Units are likely to be cm, mm, or inches.

Qualifiers: Date of measurement is important, but timing is not. The date is generally recorded with the measurement; hence, no further recommendations were necessary. It was consequently decided that qualifiers should be optional for some data items.

Value ranges: Most clinicians among the attendees felt that using value ranges within the paediatric data dictionary could be problematic due to 1) the need for different value ranges for different paediatric age groups, 2) the study protocol providing definitions, and 3) children routinely presenting with values outside of the normal range. It would then be necessary to consider what action might be triggered by an entry outside of this range. For example, would the system block such an entry? Would an alert trigger? These questions require further consensus and could perhaps be a part of the second iteration of the CCPDD. Hence, it was decided not to consider value ranges for data items at this workshop.

---

Three items—fontanelle closure, pregnancy-related events and stool sampling—were determined to be out of scope and were rejected at the vote. They were contentious items at the preceding discussions as well. Fontanelle closure was seen as very specific for some neonate diseases, but the group mostly agreed this is not cross cutting. Similarly, it was agreed that stool samples are not collected frequently enough across studies to be considered cross cutting. For pregnancy-related events, the major discussion revolved around distinguishing between maternal pregnancy and items relating to an adolescent girl who may be pregnant. While maternal pregnancy items may be cross cutting (e.g., gestational age), pregnancy-related events for an adolescent child were deemed to be out of scope.

Six items—infant feeding, development, APGAR score, concomitant medications, comorbidities and standard blood tests—were accepted during the vote but later deferred to the next iteration of the CCPDD. The main reasoning was that further evaluation regarding paediatric specificity of these items was necessary. For infant feeding, the attendees agreed that "breastfed or not" was within scope but would be incomplete without length and level (fully, partial, none) of feeding. These required precise definitions. Measuring child development is cross cutting but is a huge area including functional, social and motor development. Several sub-items including questionnaires ratings and scales would need to be evaluated, including whether existing CDISC standards covered these topics. CDISC standards representing data from the AGPAR score were currently in development at the time of the workshop. While it was debated whether certain common blood tests—creatinine, electrolytes, glucose and complete blood count—should be included in this version of the CCPDD, it was agreed to add the full set of blood tests to the next CCPDD iteration. Further feedback was sought from CDISC about the addition of concomitant medications.

**Table 1:** Results from the vote for inclusion of data items in the CCPDD. Colour codes: Green – Majority vote for inclusion and subsequently included; Red – Majority vote for exclusion due to item being out of scope; Yellow – Majority vote for inclusion but deferment or further review required.

| Item | Yes % | No % | Deferred | Notes |
|---|---|---|---|---|
| Date of birth | 100 | 0 | | |
| Estimated gestational age | 100 | 0 | | |
| Phenotypic Sex/Sex | 100 | 0 | | Questions remained whether to use sex or phenotypic sex |
| Date of death | 100 | 0 | | |
| Head circumference | 100 | 0 | | |
| Blood pressure | 100 | 0 | | Systolic and diastolic blood pressure |
| Body mass index | | | | Replaced by body surface area |
| Body surface area | | | | Replaced item – no voting |
| Height | 100 | 0 | | |
| Weight | 100 | 0 | | |
| Total body length | 91.7 | 8.3 | | |
| Heart rate | 100 | 0 | | |
| Pulse rate | 81.8 | 18.2 | | |
| Respiratory rate | 75 | 25 | | |
| Oxygen saturation | 81.8 | 18.2 | | |
| Temperature | 83.3 | 16.7 | | |
| Diagnosis | 91.7 | 8.3 | | |
| Fontanelle closure | 36.3 | 63.6 | | |
| Pubertal status | 75 | 25 | | Split into Tanner Staging (stages for male genitalia, male pubic hair, female breast, female pubic hair), testicular volume and menarche (date and age) |
| Pregnancy related events | 18.2 | 81.8 | | |
| Infant feeding | 70 | 30 | X | |
| Stool sampling | 44.5 | 54.5 | | |
| Development | 100 | 0 | X | |
| APGAR score | 100 | 0 | X | |
| Concomitant medications | 92.3 | 7.7 | X | |
| Comorbidities | 84.6 | 15.4 | X | |
| Standard blood tests | 100 | 0 | X | |

A final decision at the workshop was to split pubertal stage into three subcategories: Tanner staging[28] (stages for male genitalia, male pubic hair, female breast, female pubic hair), testicular volume and menarche (date and age). One point of debate was whether testicular volume should be separate from Tanner staging.

The final changes were applied to the CCPDD after review of the included items from CDISC following the workshop. The removal of BMI was reversed, and both BMI and BSA were retained. The rationale behind this decision was that while BSA was more relevant for younger patients, BMI was a routinely collected data item for older paediatric patients. The term phenotypic sex was agreed to be included in the CCPDD and would be discussed during the development of the CDISC PUG in terms of updating the CDISC definition of sex. The addition of "diagnoses" was pushed back to the next iteration as it needed proper differentiation from "comorbidities." Concomitant medications, which had previously been pushed back to the next iteration, was brought forward into the first version as current CDISC modelling of concomitant medications data seemed sufficient. The final dictionary with 25 terms was released on August 29, 2019, as a Microsoft Excel file. While sharing the entire dictionary is prohibited (reasons explained below), two snapshots are shared in **Figures 2** and **3**.

| Sub-topic | CDISC mapped Item | CDISC units (CDISC Codelist) C-CODE value in brackets | CDISC SDTM Domain | CDISC Qualifier Variables | Guidance | Source of information |
|---|---|---|---|---|---|---|
| Tanner staging | Male Genitalia Stage | N/A | Questionnaires, Ratings, and Scales (QS) | | Tanner Staging is modelled in the CDISC Therapeutic Area Data Standards for Type 1 Diabetes - Pediatrics and Devices Modules | https://www.cdisc.org/system/files/members/standard/ta/TAUG-T1D_Pediatrics_and_Devices_v2.0.pdf |
| | Male Pubic Hair Stage | N/A | Questionnaires, Ratings, and Scales (QS) | | Tanner Staging is modelled in the CDISC Therapeutic Area Data Standards for Type 1 Diabetes - Pediatrics and Devices Modules | https://www.cdisc.org/system/files/members/standard/ta/TAUG-T1D_Pediatrics_and_Devices_v2.0.pdf |
| | Female Breast Stage | N/A | Questionnaires, Ratings, and Scales (QS) | | Tanner Staging is modelled in the CDISC Therapeutic Area Data Standards for Type 1 Diabetes - Pediatrics and Devices Modules | https://www.cdisc.org/system/files/members/standard/ta/TAUG-T1D_Pediatrics_and_Devices_v2.0.pdf |
| | Female Pubic Hair Stage | N/A | Questionnaires, Ratings, and Scales (QS) | | Tanner Staging is modelled in the CDISC Therapeutic Area Data Standards for Type 1 Diabetes - Pediatrics and Devices Modules | https://www.cdisc.org/system/files/members/standard/ta/TAUG-T1D_Pediatrics_and_Devices_v2.0.pdf |
| Menarche | Date of menarche | | Medical History (MH) Clinical Events (CE) | | See the CDISC Therapeutic Area Data Standards for Type 1 Diabetes - Pediatrics and Devices Modules  Refer to the followng CDISC documentation on recommendations on how to collect dates/times in | https://www.cdisc.org/system/files/members/standard/ta/TAUG-T1D_Pediatrics_and_Devices_v2.0.pdf |
| | Age of menarche | | Reprodictive System Findings (RP) | | See the CDISC Therapeutic Area Data Standards for Type 1 Diabetes - Pediatrics and Devices Modules | https://www.cdisc.org/system/files/members/standard/ta/TAUG-T1D_Pediatrics_and_Devices_v2.0.pdf |
| Testicular Volume | Testicular Volume | UNIT Codelist (C71620) mL (C28254) | Reprodictive System Findings (RP) | Location (RPLOC) = TESTES Laterality (RPLAT) = LEFT or RIGHT Method (RPMETHOD) = e.g., ORCHIDOMETRY | See the CDISC Therapeutic Area Data Standards for Type 1 Diabetes - Pediatrics and Devices Modules | https://www.cdisc.org/system/files/members/standard/ta/TAUG-T1D_Pediatrics_and_Devices_v2.0.pdf |

**Figure 2:** A snapshot of the pubertal status section of the cross-cutting paediatric data dictionary.

| CDISC mapped Item | CDISC units (CDISC Codelist) C-CODE value in brackets | CDISC SDTM Domain | CDISC Qualifier Variables | Guidance | Source of information CDASH = Data Collection SDTM = Data Tabulation |
|---|---|---|---|---|---|
| Height | cm (C49668) m (C41139) in (C48500) ft (C71253) | Vital Signs (VS) | unit, standardised unit | Specify standardised unit | CDASHIG v2.1 - https://www.cdisc.org/system/files/members/standard/foundational/CDASHIG%20v2.1-Final_Rev.pdf Section 8.3.18 SDTMIG v3.3 - https://www.cdisc.org/standards/foundational/sdtmig/sdtmig-v3-3/html#Vital+Signs |
| Total body Length | cm (C49668) in (C48500) ft (C71253) | Vital Signs (VS) | Vital Signs Position of Subject (VSPOS)? - not needed if new definition accepted as it states "recumbent" | Specify standardised unit | CDASHIG v2.1 - https://www.cdisc.org/system/files/members/standard/foundational/CDASHIG%20v2.1-Final_Rev.pdf Section 8.3.18 SDTMIG v3.3 - https://www.cdisc.org/standards/foundational/sdtmig/sdtmig-v3-3/html#Vital+Signs |
| Weight | kg (C28252) g (C48155) LB (C48531) No CDISC controlled terminology exists for stone | Vital Signs (VS) | | Specify standardised unit | CDASHIG v2.1 - https://www.cdisc.org/system/files/members/standard/foundational/CDASHIG%20v2.1-Final_Rev.pdf Section 8.3.18 SDTMIG v3.3 - https://www.cdisc.org/standards/foundational/sdtmig/sdtmig-v3-3/html#Vital+Signs |
| BMI | kg/m2 (C49671) | Vital Signs (VS) | | Raw data values must be collected (Height and weight) | CDASHIG v2.1 - https://www.cdisc.org/system/files/members/standard/foundational/CDASHIG%20v2.1-Final_Rev.pdf Section 8.3.18 SDTMIG v3.3 - https://www.cdisc.org/standards/foundational/sdtmig/sdtmig-v3-3/html#Vital+Signs |
| Body surface area | m2 (C42569) | Vital Signs (VS) | Method used: Haycock Gehan and George Boyd Fujimoto Shuter and Aslani Schlich | Need to define the options - which are used/not used in paediatrics | CDASHIG v2.1 - https://www.cdisc.org/system/files/members/standard/foundational/CDASHIG%20v2.1-Final_Rev.pdf Section 8.3.18 SDTMIG v3.3 - https://www.cdisc.org/standards/foundational/sdtmig/sdtmig-v3-3/html#Vital+Signs |

**Figure 3:** A snapshot of the vital signs section of the cross-cutting paediatric data dictionary. Text in red may be amended in version 2 of the dictionary.

## Discussion

Immediately upon release, the CCPDD was accepted as a tool for creating CRFs in three proof of viability paediatric clinical trials: A New Posaconazole Dosing Regimen for Paediatric Patients With Cystic Fibrosis and Aspergillus Infection (cASPerCF, ClinicalTrials.gov Identifier: NCT04966234, EudraCT Number: 2019-004511-31), Kawasaki Disease Coronary Artery Aneurysm Prevention trial (KD-CAAP, EudraCT Number: 2019-004433-17) and Prophylactic Treatment of the Ductus Arteriosus in Preterm Infants by Acetaminophen (TREOCAPA – ClinicalTrials.gov Identifier: NCT04459117, EudraCT Number: 2019-004297-26). Four more industry-sponsored trials were likely to commence in 2021 but were delayed due to the COVID-19 pandemic and pharmaceutical companies staggered "first patient first visit" timelines.

The inclusion of CDISC within the development process ensured that the CCPDD contained CDISC implementation notes and examples, and therefore the CCPDD was accessible immediately upon release (without requiring any format conversions and translations). The standard procedures used for collection of the cross-cutting data items will enable pooled data from different sources being interoperable and promises their combined reuse in future research. Such interoperability of data is particularly important for a field like paediatrics where regular research methodologies are constrained by limited populations, regulatory challenges, and ethical considerations. Hence, the long-term use of the data dictionary will facilitate the interoperability and reusability aspects of the FAIR principles. The findability and accessibility aspects of the data will remain the responsibility of the data provider.

While cross-cutting data items were an appropriate starting point, the major challenge will be extending the dictionary beyond cross-cutting terms. To this end, CDISC is currently developing the PUG that aims to cover examples of data collection and tabulation for paediatric terms used in clinical trials. The CCPDD is one of the inputs for the PUG, but other inputs will also be used. The PUG is in its final stages of public review and expected to be available on the CDISC website by the end of 2022.[29]

Though a rigorous process was used for the creation of the CCPDD, the dictionary had certain limitations. The very nature of consensus-building meant that a different group of experts may have arrived at a different set of data items. While every effort was made to achieve unanimous consensus, certain decisions did require voting, for example, the vote on stool sampling. The development in Microsoft Excel led to a flat structure of the CCPDD with poor visualization. Hence, c4c is exploring whether the next iteration of the CCPDD could be based on a clinical modelling tool (CMT).[30] Such a tool has already been piloted during the project and includes mind-map visualizations as well as export to well-known data formats (e.g., JSON, XML).

The above-mentioned next iteration of the CCPDD is currently under development. Along with the additional data items and possibly the CMT, this iteration will include a change in control process, which will allow for new items of relevance to be added continuously. In addition, there are several other strands of c4c work that are likely to expand the CCPDD in future. These include exploring which real-world data items are most relevant in paediatric phase IV (post-marketing surveillance) trials and reuse of data for comparator arms in rare disease studies.[20]

## Conclusion

c4c is creating the CCPDD to meet a need in the field of paediatric research: by increasing the standardization of data collected within paediatric clinical trials. The team used a highly collaborative, consensus building methodology to ensure the most relevant data items were included in the dictionary. The CCPDD focusses on items that are both commonly collected and cross cutting, maximising the potential impact of its implementation.

## Disclaimer

The publication reflects the authors' view and neither IMI nor the European Union, EFPIA, or any Associated Partners are responsible for any use that may be made of the information contained therein.

## Acknowledgements

This project has received funding from the Innovative Medicines Initiative 2 Joint Undertaking under grant agreement no. 777389. The Joint Undertaking receives support from the European Union's Horizon 2020 research and innovation programme and EFPIA.

## Competing Interests

The authors have no competing interests to declare.

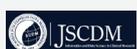

*Journal of the Society for Clinical Data Management* is a peer-reviewed open access journal published by Society for Clinical Data Management.

OPEN ACCESS